\documentclass{PoS}

\title{NLO exclusive diffractive processes with saturation}

\ShortTitle{NLO exclusive diffractive processes with saturation}

\author{\speaker{R.~Boussarie}\\
        Institute of Nuclear Physics, Polish Academy of Sciences, Radzikowskiego 152, PL-31-342 Krak\'{o}w, Poland\\
        E-mail: \email{renaud.boussarie@ifj.edu.pl}}
        
\author{A.~V.~Grabovsky\\
        Novosibirsk State University, 2 Pirogova street, 630090 Novosibirsk, Russia and Budker Institute of Nuclear Physics, 11 Lavrenteva avenue, 630090 Novosibirsk, Russia\\
        E-mail: \email{a.v.grabovsky@inp.nsk.su}}

\author{D.~Yu.~Ivanov\\
        Novosibirsk State University, 2 Pirogova street, 630090 Novosibirsk, Russia and Sobolev Institute of Mathematics, 630090 Novosibirsk, Russia\\
        E-mail: \email{d-ivanov@math.nsc.ru}}

\author{L.~Szymanowski\\
        National Centre for Nuclear Research (NCBJ), 00-681 Warsaw, Poland\\
        E-mail: \email{lech.szymanowski@ncbj.gov.pl}}

\author{S.~Wallon\\
        Laboratoire de Physique Th\'eorique (UMR 8627), CNRS, Univ. Paris-Sud, CNRS, Universit\'{e} Paris-Saclay, 91405 Orsay Cedex, France and UPMC Univ. Paris 06, Facult\'{e} de Physique, 4 place Jussieu, 75252 Paris Cedex 05, France\\
        E-mail: \email{samuel.wallon@th.u-psud.fr}}

\abstract{We present two NLO exclusive impact factors computed in the QCD shock wave approach. These are the very first steps towards precision studies of a wide range of high energy exclusive processes with saturation effects in $ep$, $eA$, $pp$ and $pA$ collisions.}

\FullConference{XXV International Workshop on Deep-Inelastic Scattering and Related Subjects\\
		3-7 April 2017\\
		University of Birmingham, UK}

\begin{document}

\section{Introduction}

We will present two results, published in refs.~\cite{Boussarie:2014lxa}, \cite{Boussarie:2016ogo} and \cite{Boussarie:2016bkq}, for one-loop computations of impact factors for high-energy diffractive DIS-like exclusive processes. Such results, once numerically solved, would provide the first complete Next-to-Leading-Logarithmic description of many exclusive processes with or without saturation. The impact factors obtained here rely on the QCD shock wave formalism~\cite{Balitsky:1995ub-Balitsky:1998kc-Balitsky:1998ya-Balitsky:2001re}, equivalent to the color-glass condensate framework~\cite{JalilianMarian:1997jx-JalilianMarian:1997gr-JalilianMarian:1997dw-JalilianMarian:1998cb-Kovner:2000pt-Weigert:2000gi-Iancu:2000hn-Iancu:2001ad-Ferreiro:2001qy}, which extends the BFKL formalism by including gluonic saturation effects. In this framework, we will describe the \textit{direct} Pomeron contribution to diffraction as the action of color singlet Wilson line operators on the target fields.
We will put an emphasis on the various mechanisms for the cancellation of divergences in both processes considered.

\section{The shockwave formalism}
We will work in $D \equiv d+2 \equiv 4+2\epsilon$ dimensions. We introduce two lightcone vectors $n_1$ and $n_2$ such that the projectile (resp. target) flies mainly along $n_1$ (resp. $n_2$). They define a Sudakov basis, so for any $D$-vectors $k$ and $l$ we will write
\begin{equation}
\label{Sudakov}
k\equiv k^+ n_1 + k^- n_2 + k_\perp, \quad k\cdot l = k^+ l^- + k^- l^+ - \vec{k}\cdot\vec{l}.
\end{equation}
We use the $n_2$ lightcone gauge for gluonic fields $\mathcal{A}\cdot n_{2}=0,$
and write $\mathcal{A}$ as the sum of an external field $b_\eta$ built from slow gluons 
whose momenta are limited by the longitudinal cutoff $e^\eta p_\gamma^+$, where $\eta$ is an arbitrary negative parameter and $p_\gamma$ is the momentum of the incoming projectile, and the remaining quantum field $A_\eta$. In the high center-of-mass energy limit considered here, we can write
\begin{equation}
\mathcal{A}^{\mu} = A^{\mu}_\eta+b_\eta^{\mu},\;\;\;\;\;\;\;\;\;\quad b_\eta^{\mu}\left(  z\right)  =b_\eta^{-}(z^{+},\vec{z}\,) \,n_{2}%
^{\mu}=\delta(z^{+})B_\eta\left(  \vec{z}\,\right)  n_{2}^{\mu}\,,\label{b}%
\end{equation}
where
$B_\eta(\vec{z})$ is a profile function.
We then define path-ordered Wilson lines as 
\begin{equation}
U_{\vec{z}_{i}}^\eta = \mathcal{P} \exp\left[{ig\int_{-\infty
}^{+\infty}b_{\eta}^{-}(z_{i}^{+},\vec{z}_{i}) \, dz_{i}^{+}}\right]\,.
\label{WL}%
\end{equation}
From the Wilson lines, we define the dipole operator and its Fourier transforms as follows:
\begin{eqnarray}
& & \mathcal{U}_{\vec{z}_i\vec{z}_j}^\eta \equiv \mathrm{Tr}(U_{\vec{z}_i}^\eta U_{\vec{z}_j}^{\eta\dagger}) - N_c, \\
& & \tilde{\mathcal{U}}_{\vec{p}_i\vec{p}_j}^\eta \equiv \! \int \! d^d\vec{z}_i \, d^d\vec{z}_j \, e^{-i(\vec{p}_i \cdot \vec{z}_i) -i(\vec{p}_j \cdot \vec{z}_j)} \mathcal{U}_{\vec{z}_i\vec{z}_j}^\eta, \\
& & \tilde{\mathcal{V}}_{\vec{p}_i\vec{p}_j\vec{p}_k}^\eta \equiv \! \int \! d^d\vec{z}_i \, d^d\vec{z}_j \, d^d\vec{z}_k e^{-i(\vec{p}_i \cdot \vec{z}_i) -i(\vec{p}_j \cdot \vec{z}_j) -i(\vec{p}_k \cdot \vec{z}_k)} \mathcal{U}^\eta_{\vec{z}_i\vec{z}_k}\mathcal{U}^\eta_{\vec{z}_k\vec{z}_j}.
\end{eqnarray}
When computing a physical amplitude, one acts with such operators on the incoming and outgoing states of the target. For example in the case of a coherent diffractive $\gamma^{(\ast)}(p_\gamma) P(p_0) \rightarrow X(p_X) P^\prime(p_0^\prime)$ process, the following matrix elements will be involved:
\begin{equation}
\mathcal{W}^\eta \rightarrow \langle P^\prime (p_0^\prime) \vert \mathcal{P}(\mathcal{W}^\eta) \vert P(p_0) \rangle,
\end{equation}
where $\mathcal{W}^\eta$ is an operator built from the Wilson lines. In both diffractive cases considered here, there are two possibilities for $\mathcal{W}^\eta$: either a dipole operator $\mathcal{W}^\eta = \tilde{\mathcal{U}}_{\vec{p}_i\vec{p}_j}^\eta$, or a double-dipole operator $\mathcal{W}^\eta = \tilde{\mathcal{V}}_{\vec{p}_i\vec{p}_j\vec{p}_k}^\eta$. Note that in the t'Hooft limit $N_c^{-2} \rightarrow 0$ or in the mean field approximation, the matrix elements for the double dipole operators can be written as the product of the matrix elements for two dipole operators. From now on we will write $\mathcal{W}$ rather than $\mathcal{W}^\eta$ for readability.

\section{Impact factors for open $q\bar{q}$ and open $q\bar{q}g$ production}

\subsection{Impact factor for open $q\bar{q}$ production}

\begin{figure}
\center
\includegraphics[scale=0.35]{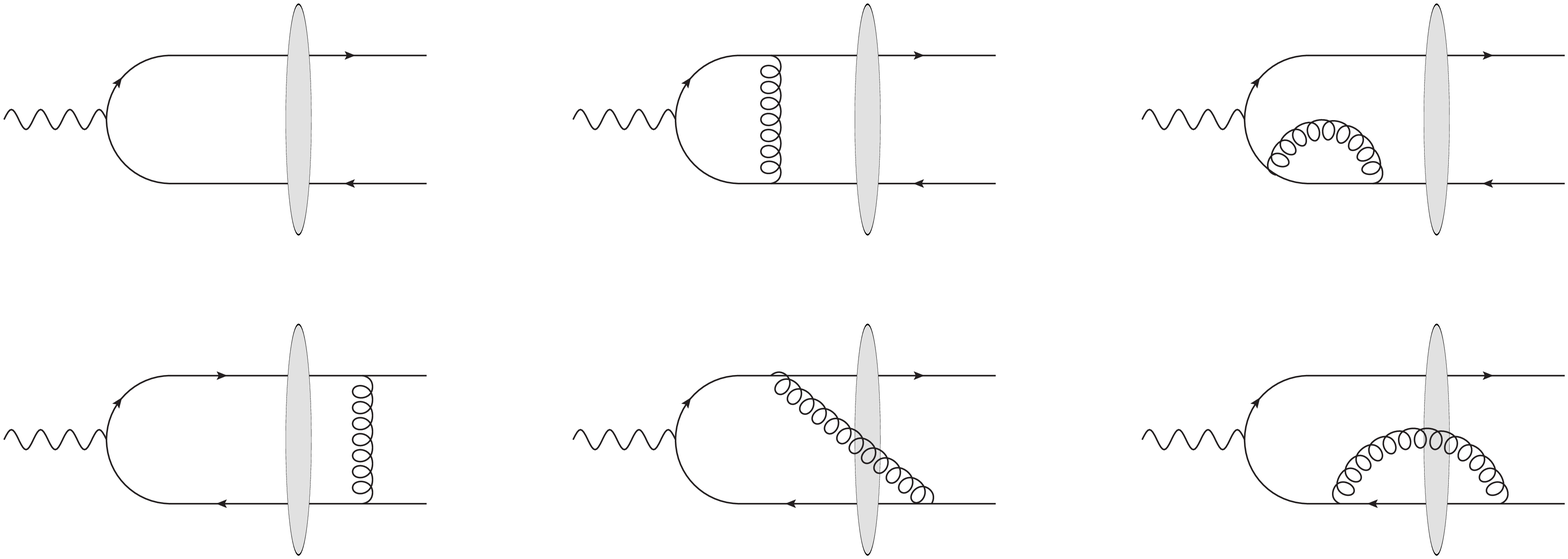}
\caption{Diagrams for the impact factor for open $q\bar{q}$ production. The gray blob stands for the interaction with the external shock wave field via the convolution with the path-ordered Wilson line operators}
\label{diagrams}
\end{figure}

The diagrams contributing to the impact factor for the $\gamma^{\ast}\rightarrow q\bar{q}$ transition are shown in figure~\ref{diagrams}.
After the projection on the color singlet state and the subtraction of the contribution without interaction with the external field, the contribution of these diagrams can be written in the momentum space as the convolution of Wilson line operators with what we will refer to as impact factors with the following form:
\begin{eqnarray}
& & M^{q\bar{q}} = \varepsilon_\mu \int \! d^d\vec{p}_1 \, d^d\vec{p}_2 \, d^d\vec{p}_3 \, \delta(\vec{p}_{q1} + \vec{p}_{\bar{q}2}-\vec{p}_3) \delta(p_q^++p_{\bar{q}}^+-p_\gamma^+) \label{virtualconv} \\
 & \times & \left\{  \tilde{\mathcal{U}}_{\vec{p}_1\vec{p}_2} \, \delta(\vec{p}_3)\left[ \Phi_0^\mu +  C_F \left( \Phi_{V_1}^\mu + \Phi_{V_2}^\mu \right) \right] \nonumber + \, C_A \left( \tilde{\mathcal{V}}_{\vec{p}_1\vec{p}_2\vec{p}_3} + \tilde{\mathcal{U}}_{\vec{p}_1\vec{p}_3} + \tilde{\mathcal{U}}_{\vec{p}_3\vec{p}_2} - \tilde{\mathcal{U}}_{\vec{p}_1\vec{p}_2} \right) \Phi_{V_2}^\mu\right\}, \nonumber
\end{eqnarray} 
where we denoted $p_{ij} \equiv p_i-p_j$, and where $p_q$ (resp. $p_{\bar{q}}$) is the momentum of the outgoing quark (resp. antiquark). $\Phi_0^\mu = \Phi_0^\mu(\vec{p}_1,\,\vec{p}_2)$ is directly obtained by computing the first diagram in figure~\ref{diagrams}, $\Phi_{V_1}^\mu = \Phi_{V_1}^\mu(\vec{p}_1,\,\vec{p}_2)$ is obtained from the second, third and fourth diagram\footnote{Note that the contributions of the first two diagrams in $\Phi_{V_1}$ were also obtained in ref.~\cite{Beuf:2016wdz}.} and $\Phi_{V_2}^\mu = \Phi_{V_2}^\mu(\vec{p}_1,\,\vec{p}_2,\,\vec{p}_3)$ is obtained from the last two diagrams.

Several divergences appear in each of the NLO terms in eq.~(\ref{virtualconv}): $\Phi_{V_1}^\mu$ contains soft, collinear, soft and collinear, and UV divergences, while $\Phi_{V_2}^\mu$ contains a rapidity divergence\footnote{i.e. a divergence for $k^+ \rightarrow 0$ for fixed $k^-$ and $k_\perp$, due to the spurious lightcone gauge pole.}. In the shockwave formalism and in lightcone gauge, it is impossible to use the usual dimensional regularization around dimension 4 due to the presence of the cutoff on $p^+$ momenta: the 2 longitudinal directions must be isolated. Thus we use dimensional regularization $d=2+2\epsilon$ for the transverse components, and the cutoff prescription $k^+ < e^\eta p_\gamma^+$ which is natural in our formalism.   

The rapidity divergence in $\Phi_{V_2}^\mu$ is canceled via the use of the B-JIMWLK evolution equation for the dipole operator: evolving the dipole operator in the leading order convolution in~(\ref{virtualconv}) w.r.t. the longitudinal cutoff from the arbitrary $e^\eta p_\gamma^+$ to a more physical divide $e^{\eta_0} p_\gamma^+$, which will serve as a factorization scale which separates the upper and lower impact factors, allows one to cancel the dependence on $\eta$ in $\Phi_{V_2}^\mu$ and get a finite expression for the double-dipole contribution to the NLO impact factor. In momentum space and in $d+2$ dimensions, the evolution equation is given by:
\begin{eqnarray}
& & \hspace*{-.32cm} \frac{\partial{\tilde{\mathcal{U}}_{\vec{p}_1\vec{p}_2}^{\eta}}}{\partial\mathrm{log}\eta} = 2\alpha_{s}\mu^{2-d}\!\!\!\int\!\frac{d^{d}\vec{k}_{1}d^{d}\vec{k}_{2}d^{d}\vec{k}_{3}}{\left(2\pi\right)^{2d}}\delta\!\left(\vec{k}_{1}+\vec{k}_{2}+\vec{k}_{3}-\vec{p}_{1}-\vec{p}_{2}\right)\!\!\!\left(\!\tilde{\mathcal{V}}_{\vec{k}_1\vec{k}_2\vec{k}_3}^\eta\!+ N_c\!\left(\!\tilde{\mathcal{U}}_{\vec{k}_1\vec{k}_3}^\eta\!+\tilde{\mathcal{U}}_{\vec{k}_3\vec{k}_2}^\eta\!-\tilde{\mathcal{U}}_{\vec{k}_1\vec{k}_2}^\eta\right)\!\!\right) \!\!\!\nonumber \\
 &  & \hspace*{-.3cm} \times  \left[2\frac{(\vec{k}_{1}-\vec{p}_{1}).(\vec{k}_{2}-\vec{p}_{2})}{(\vec{k}_{1}-\vec{p}_{1})^{2}(\vec{k}_{2}-\vec{p}_{2})^{2}}+\frac{\pi^{\frac{d}{2}}\Gamma\left(1-\frac{d}{2}\right)\Gamma^{2}\left(\frac{d}{2}\right)}{\Gamma\left(d-1\right)}\left(\frac{\delta(\vec{k}_{2}-\vec{p}_{2})}{\left[(\vec{k}_{1}-\vec{p}_{1})^{2}\right]^{1-\frac{d}{2}}}+\frac{\delta(\vec{k}_{1}-\vec{p}_{1})}{\left[(\vec{k}_{2}-\vec{p}_{2})^{2}\right]^{1-\frac{d}{2}}}\right)\right]\!\!. \label{JIMWLK}
\end{eqnarray}
Evolving the Wilson lines from the arbitrary cutoff $\eta$ to the rapidity divide $\eta_0$, which has the role of a $t$-channel factorization scale in the shockwave framework, creates a counterterm to $\Phi_{V2}^\mu$ which reads:
\begin{equation}
\label{BKcounterterm}
\tilde{\Phi}_{V2}^\mu = -\frac{\mu^{2-d}}{\Gamma(1-\epsilon)\pi^{1+\epsilon}}\ln\left(\frac{e^{\eta_0}}{e^{\eta}}\right)\int d^d\vec{k}_1 d^d\vec{k}_2 \delta(\vec{p}_1+\vec{p}_2+\vec{p}_3-\vec{k}_1-\vec{k}_2) \mathcal{H}(\vec{k}_1,\,\vec{k}_2,\,\vec{k}_3,\,\vec{p}_1,\vec{p}_2),
\end{equation}
where $\mathcal{H}$ is the $d$-dimensional Balitsky-Kovchegov kernel\footnote{At $d=2$, see refs.~\cite{Balitsky:1995ub-Balitsky:1998kc-Balitsky:1998ya-Balitsky:2001re,Kovchegov:1999yj-Kovchegov:1999ua}.} in momentum space and can be extracted from eq.~(\ref{JIMWLK}).
Adding this counterterm allows one to get rid of the rapidity divergence. A similar countertem arises when varying the rapidity divide, and a similar cancellation mechanism then allows one to cancel the dependence on that scale up to Next-to-Next-to-Leading-Logarithmic (NNLL) corrections.

The divergences in $\Phi_{V_1}$ must be canceled in a process-dependent way. In the following sections we will show how to cancel them at the level of the amplitude for a process and at the level of the cross-section for a second process.

\subsection{Impact factor for open $q\bar{q}g$ production}

\begin{figure}
\center
\includegraphics[scale=0.5]{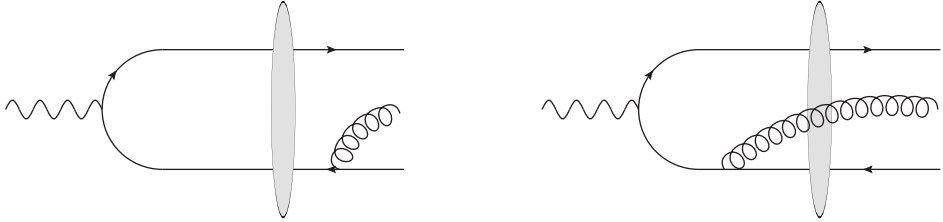}
\caption{Diagrams contributing to the impact factor for open $q\bar{q}g$ production.}
\label{3body}
\end{figure}

The convolution for the $\gamma^{(\ast)}\rightarrow q\bar{q}g$ impact factor is very similar to the one for the $\gamma^{(\ast)}\rightarrow q\bar{q}$ impact factor:
\begin{eqnarray}
M^{q\bar{q}g} & = & \varepsilon_\mu \int \! d^d\vec{p}_1 \, d^d\vec{p}_2 \, d^d\vec{p}_3 \, \delta(\vec{p}_{q1} + \vec{p}_{\bar{q}2}+\vec{p}_{g3}) \delta(p_q^++p_{\bar{q}}^++p_g^+-p_\gamma^+) \label{realconv} \\	
 & \times & \left\{ C_F \left( \Phi_{R_1}^\mu + \Phi_{R_2}^\mu \right) \tilde{\mathcal{U}}_{\vec{p}_1\vec{p}_2} \, \delta(\vec{p}_3) + \, C_A  \left( \tilde{\mathcal{V}}_{\vec{p}_1\vec{p}_2\vec{p}_3} + \tilde{\mathcal{U}}_{\vec{p}_1\vec{p}_3} + \tilde{\mathcal{U}}_{\vec{p}_3\vec{p}_2} - \tilde{\mathcal{U}}_{\vec{p}_1\vec{p}_2} \right) \Phi_{R_2}^\mu \right\}, \nonumber
\end{eqnarray} 
where $\Phi_{R_1}^\mu=\Phi_{R_1}^\mu(\vec{p}_1, \, \vec{p}_2)$ and $\Phi_{R_2}^\mu=\Phi_{R_2}^\mu(\vec{p}_1, \, \vec{p}_2,\,\vec{p}_3)$ are obtained by computing respectively the first diagram and the second diagram in figure~\ref{3body}, as described in refs.~\cite{Beuf:2011xd}, \cite{Boussarie:2014lxa} and \cite{Ayala:2016lhd}. \\

When considering exclusive observables, the real contributions are those where the additional gluon is either collinear to the quark or to the antiquark, or too soft to be detected i.e. with an energy which is lower than a typical energy resolution $E$. The contribution from the soft gluon to the open $q\bar{q}$ cross section can be written with a very simple form:
\begin{equation}
d\sigma_{soft}^{q\bar{q}g} =  \alpha_s \left(\frac{N_c^2-1}{2N_c}\right)\int \frac{dp_g^+}{p_g^+}\frac{d^d\vec{p}_g}{(2\pi)^d} \left\vert \frac{p_q}{(p_q \cdot p_g)} - \frac{p_{\bar{q}}}{(p_{\bar{q}} \cdot p_g)} \right\vert ^2 d\sigma_{LO} , \label{RealSoft}
\end{equation}
where the integration is performed in the $p_g$-phase space region where $p_g^+ + \frac{\vec{p}_g^2}{p_g^+} < 2E$. 

In the case of vector meson production, the collinear gluon contributions cancel. We will study the collinear contribution for the dijet case in section~\ref{sec:impact-dijet}.
All the infrared divergences in the $q\bar{q}g$ contribution will be combined in a process-dependent way to cancel the remaining divergences in the virtual terms.

\section{Impact factor for the production of a longitudinally polarized light vector meson}
\label{sec:impact-meson}

We will now build a finite amplitude for the production of a light vector meson $V$ (e.g. $V = \rho, \phi, \omega$). For this purpose, in addition to the CGC rapidity separation in $t$-channel we will use leading-twist collinear factorization in $s$-channel. Let us define the twist-2 Distribution Amplitude (DA) for the longitudinally polarized meson via the expansion of the vacuum-to-meson matrix element of the leading twist 2-particle operator:
\begin{equation}
\label{DA}
<V_L(p_V)|\bar{\psi}(z)\gamma^\mu \psi(0)|0>_{z^2 \rightarrow 0} \, = \, f_V p_V^\mu \int_0^1 dx \, e^{ix(p_V \cdot z)} \varphi(x,\mu_F).
\end{equation}
To obtain the hard matrix element for this process, one only needs to substitute 
\begin{equation}
(p_q,p_{\bar{q}}) \rightarrow (x p_V,\bar{x}p_V), \quad (\bar{u}_{p_q})_{\alpha}(v_{p_{\bar{q}}})_{\beta} \rightarrow \frac{1}{4} \gamma^\mu_{\beta\alpha} \label{ColKin}
\end{equation}
in the result obtained from the purely diagrammatic computation and multiplying by the r.h.s. of eq.~(\ref{DA}).
Because the process is exclusive, the real correction does not contribute in the present case.
In this example of a process, the cancellation of divergences in the virtual corrections occurs at the level of the amplitude, through the Efremov-Radyushkin-Brodsky-Lepage evolution equation~\cite{Farrar:1979aw-Lepage:1979zb-Efremov:1979qk} for the DA $\varphi$, which appears when one renormalizes the bilocal operator in the r.h.s. of eq.~(\ref{DA}). In the $\overline{MS}$ renormalization scheme, it reads:
\begin{equation}
\label{ERBLevol}
\frac{\partial \varphi(x,\mu_F)}{\partial\ln\mu_F^2} = \frac{\alpha_s C_F}{2\pi} \frac{\Gamma(1-\epsilon)}{(4\pi)^\epsilon} \left(\frac{\mu_F^2}{\mu^2}\right)^{\epsilon} \int_0^1\!dz\,\varphi(z,\mu_F)\mathcal{K}(x,z),
\end{equation}
where $\mathcal{K}$ is the well-known ERBL evolution kernel
\begin{equation}
\label{ERBLkernel}
\mathcal{K}(x,z) = \frac{1-x}{1-z} \left( 1 + \left[ \frac{1}{x-z} \right]_+ \right) \theta(x-z) + \frac{x}{z} \left( 1+\left[\frac{1}{z-x}\right]_+ \right) \theta(z-x) + \frac{3}{2}\delta(x-z).
\end{equation}
Evolving the DA up to the factorization scale $\mu_F$ creates a counterterm to $\Phi_{V1}$ reading
\begin{equation}
\tilde{\Phi}_{V1}^\mu(x,\mu_F) = -\int_0^1dz\mathcal{K}(z,x)\left[\frac{1}{\epsilon} + \ln\left(\frac{\mu_F^2}{\mu^2}\right)\right]\Phi_0^\mu.
\end{equation}
This counterterm allows one to get rid of the dimensional poles and of the dependence on $\mu$. When changing the factorization scale $\mu_F$ of $\delta\mu_F$, a similar mechanism allows one to cancel the dependence on $\delta\mu_F$ up to NNLL terms.
We thus found a finite expression for the NLO amplitude for the production of a longitudinally polarized forward light vector meson. The explicit result for the finite impact factors can be found in ref.~\cite{Boussarie:2016bkq}.
Our result, for arbitrary kinematics, remains to be compared with the result of ref.~\cite{Ivanov:2004pp} in the forward kinematics and in the usual 
$k_t-$factorization framework, this last fact making this comparison non trivial.

\section{Impact factor for the production of a forward dijet}
\label{sec:impact-dijet}

The divergences in $\Phi_{V_1}$ must be canceled by combining such terms with the associated real corrections to form a physical cross section. The first step to compute such a cross section is to use a jet algorithm in order to cancel the soft and collinear divergence from the real correction. By using the jet-cone algorithm in the small cone limit, as used in ref.~\cite{Ivanov:2012ms}, we proved that such a cancellation occurs. In practice, it amounts to redefining the integration domains in eq.~(\ref{RealSoft}) and performing a redefinition of the external momenta. For example if the gluon and the quark are collinear, they will form together a single jet of momentum $p_q+p_g$, and after the right change of variables the remaining momentum $\frac{p_q^+p_g-p_g^+p_q}{p_q^+p_g^+}$ will be inclusively integrated over, in a small collinear cone region.

The thus redefined collinear contribution has a simple form, in terms of the jet variables. For example when the gluon is collinear to the quark one gets: 
\begin{equation}
d\sigma^{(qg),\,\bar{q}} = \alpha_s \left(\frac{N_c^2-1}{2N_c}\right) N_J \, d\sigma_{LO}^{jets}, \label{RealCol}
\end{equation}
where $N_J$ is proportional to the ``number of jets in the quark'', a DGLAP-type emission kernel.

The divergence in the virtual contribution can be expressed by factorizing the leading order cross section:
\begin{equation}
d\sigma_{Vdiv}^{jets} = (N_V+N_V^\ast) \, d\sigma_{LO}^{jets}, \label{VirtualDiv}
\end{equation}
where $N_V$ is extracted from the divergent part of the virtual amplitude.
As shown in ref.~\cite{Boussarie:2016ogo}, combining all divergent terms together, i.e. adding eqs.~(\ref{VirtualDiv}), (\ref{RealSoft}), (\ref{RealCol}) and the equivalent of eq.~(\ref{RealCol}) where the gluon is collinear to the antiquark, one finally obtains a finite cross section. The lengthy finite result can be found in ref.~\cite{Boussarie:2016ogo}. 

\section{Conclusion}

In the context of high-energy diffractive DIS-like processes, we built a finite amplitude for the production of a forward longitudinally polarized light vector meson and a finite cross section for the production of a forward dijet. They were described here using the QCD shock wave formalism, which generalizes the BFKL framework by including saturation effects that are expected for very high energy collisions and for heavy ion targets. In order to get a full numerical prediction for our processes, the toughest step will be to solve the dipole B-JIMWLK evolution equation with NLO accuracy. In principle it should be solved as a function of rapidity with a non-perturbative initial condition at a typical target rapidity, then evaluated at a typical projectile rapidity\footnote{Note that the observables are independent on small changes of these rapidities up to NNLL corrections.}. The convolution of the resulting non-perturbative input with the finite impact factors presented in refs.~\cite{Boussarie:2016ogo} and \cite{Boussarie:2016bkq} would then give the very first full NLL predictions with saturation for an exclusive process, providing additional observables with respect to the inclusive cases~\cite{Beuf:2017bpd,Ducloue:2017ftk}.
 Such predictions can be made for a very wide range of $ep$ and $eA$ experiments, as well as in ultraperipheral $pp$ and $pA$ collisions if one takes the (finite) photoproduction limit of the impact factors.

 Finally, let us note that our framework allows, in the case of meson exclusive production, for an inclusion of contributions beyond twist 2, based on the framework developed in refs.~\cite{Anikin:2009hk-Anikin:2009bf-Besse:2012ia}.

\paragraph*{Acknowledgements.}
\noindent
This work is partly supported by grant No 2015/17/B/ST2/01838 from the National
Science Center in Poland, by the French grant ANR PARTONS (Grant No. ANR-12- MONU-0008-01), by the
Labex P2IO and by the Polish-French collaboration agreements Polonium and COPIN-IN2P3. A.~V.~G.
acknowledges the support of RFBR Grant No. 16-02-00888 and the Dynasty
Foundation.

\providecommand{\href}[2]{#2}\begingroup\raggedright\endgroup

\end{document}